\documentclass[11pt]{article}
\usepackage{titlesec}
\titleformat{\paragraph}[runin]{\normalfont\itshape}{\theparagraph.}{.3em}{}[.]\titlespacing{\paragraph}{0pt}{1ex plus .1ex minus .2ex}{.5em}
\usepackage{amsthm,amsmath,amssymb,bm}
\usepackage{mathabx}
\usepackage[T1]{fontenc}
\usepackage[utf8]{inputenc}
\usepackage{lmodern}
\usepackage{mathtools}
\usepackage{dsfont}
\pdfoutput=1
\usepackage[english]{babel}
\usepackage[letterpaper, hmargin=1in, top=1in, bottom=1.2in, footskip=0.6in]{geometry}
\usepackage{graphicx} % Allows including images
\usepackage{booktabs} % Allows the use of \toprule, \midrule and \bottomrule in tables
\usepackage{color}
\definecolor{aquamarine}{rgb}{0.5, 1.0, 0.83}
\definecolor{ao(english)}{rgb}{0.0, 0.5, 0.0}
\definecolor{armygreen}{rgb}{0.29, 0.33, 0.13}
\definecolor{awesome}{rgb}{1.0, 0.13, 0.32}
\definecolor{ballblue}{rgb}{0.13, 0.67, 0.8}
\definecolor{bittersweet}{rgb}{1.0, 0.44, 0.37}
\definecolor{blue}{rgb}{0.0, 0.0, 1.0}
\definecolor{brinkpink}{rgb}{0.98, 0.38, 0.5}
\definecolor{ballblue}{rgb}{0.13, 0.67, 0.8}
\definecolor{brightturquoise}{rgb}{0.03, 0.91, 0.87}
\definecolor{blue-green}{rgb}{0.0, 0.87, 0.87}
\definecolor{caribbeangreen}{rgb}{0.0, 0.8, 0.6}
\definecolor{cyan}{rgb}{0.0, 1.0, 1.0}
\definecolor{amber(sae/ece)}{rgb}{1.0, 0.49, 0.0}
\graphicspath{ {images/} }
\definecolor{vdarkred}{rgb}{0.6,0,0.2}

\definecolor{vdarkred}{rgb}{0.6,0,0.2}
\definecolor{vdarkblue}{rgb}{0,0.2,0.6}
\usepackage[pdftex, colorlinks, linkcolor=vdarkblue,citecolor=vdarkred]{hyperref}

\usepackage[utf8]{inputenc}
\usepackage{lmodern}

\author{J\"urg Fr\"ohlich\footnote{Department of Physics, ETH Zurich, Switzerland}}

\title{After the Dark Ages}

\begin{document}

\maketitle

\vspace{1em}

\begin{abstract}
After recalling some puzzles in cosmology and briefly reviewing the Friedmann-Lema\^{i}tre cosmos 
a simple unified model of the ``Dark Sector'' is described. This model involves a scalar field and a pseudo-scalar 
axion field that give rise to Dark Energy in the form of ``quintessence'' and to ``fuzzy'' Dark Matter, respectively.
Predictions of the model concerning the late-time evolution of the Universe and
possible implications for the problem of the observed Matter-Antimatter Asymmetry in the Universe are 
sketched.\\

\noindent
\textit{Dedicated to Sir Michael Berry, a much admired colleague, on the occasion of his 81$^{st}$ birthday}
\end{abstract}

\section{Introduction: The Dark Sector }\label{Intro}
I much regret that, in my scientific migrations, I have never made a close encounter with the 
``planetary system'' whose central star is Michael Berry.\footnote{Alas, I am not even closely familiar with his book on 
cosmology \cite{Berry}, undoubtedly an unforgivable gap in my education.} It is a great pleasure and honour for me 
to offer him my best wishes for a luminous future!

A more appropriate title for this little essay might be: ``After the Dark Ages is before a \mbox{Dark Age.'' --}
The early Middle Ages, after the fall of the Western Roman Empire, have been dubbed \textit{``Dark Ages''} by the 
Renaissance scholar and poet Francesco Petrarcha, who thought of the post-Roman centuries 
as ``dark,'' compared to the ``light'' era of classical antiquity. 

One cannot help worrying that, after the dark ages of two World Wars, during the first half of the past century, 
humanity presently faces the threat of a new ``dark age.'' A little more than thirty years after the post-war era of 
a bi-polar world dominated by the United States of America and the Soviet Union, which was reasonably, 
if precariously, stable and predictable, with the decline of democratic structures in some of 
the European countries and in America, and with various environmental catastrophes looming,
the general situation in the world appears to have become very unstable and fragile again. 
Recent developments in Eastern Europe are particularly frightening. 
It is important to ponder how the present dangerous situation could be changed for the better, and
how humanity may set out to enter the dawn of a future ``light era.'' This would be a worthy 
subject for an essay like this one. But I won't address it here, except for a quote at the end of this paper.

In the evolution of the cosmos, the period between the last scattering of photons from the cosmic microwave background 
(CMB) by the homogeneous plasma to the later formation of luminous structure caused by gravitational collapse is 
known as the \textit{``Dark Ages''} of the Universe; see, e.g., \cite{Jordi, Ostriker}. During this period, which lasted from 
roughly 500'000 years after the Big Bang to an age of approximately \mbox{5 $\times 10^{8}$} years, the oldest stars 
were formed. Afterwards, a period of reionization started, which was triggered by the ionizing light from the first stars
and which ended when essentially all atoms in the intergalactic medium had been re-ionized. 
According to the \textit{Cold Dark Matter} theory, structure formation in the Universe actually started much 
earlier. Apparently, it was  first caused by the presence of Dark Matter (DM) in the Universe. 
At present, DM appears to be roughly six times more abundant than visible matter and accounts for around one 
quarter of the total energy density in the Universe. The remaining somewhat more than two thirds of the energy 
density are contributed by Dark Energy (DE), which is responsible for the observed accelerated expansion
of the Universe. It is likely that it will continue to exhibit accelerated expansion to finally enter a ``Dark Age.''

The degrees of freedom of Dark Matter and Dark Energy form together what, in speculative theories, is 
called the \textit{Dark Sector}.\footnote{The name ``Dark Sector'' originates from a ``third-person shooter 
video game developed by `Digital Extremes' for the Xbox 360, PlayStation 3 and Microsoft Windows.'' 
Well, that is not exactly an essential piece of information for what we intend to describe in this essay. 
But it is apparently the origin of the name.}
The nature of the degrees of freedom that constitute the Dark Sector is unknown; hence it is made the subject of 
theoretical speculation. 
In this paper, some tentative ideas about it are discussed, and it is sketched what they tell us about
the evolution of the Universe after the Dark Ages. I hasten to warn potential readers that I 
am \textit{not} a professional cosmologist and that I cannot guarantee that the Dark-Sector model 
sketched in the following is realistic (see \cite{Fr1, BFN, BF}). But, in this dark age, it is good
to engage in somewhat insane speculations that will distract us from the insanity of the world.

I have thought about various puzzles encountered in cosmology for more than two decades (see 
\cite{FP, BFR1, BFR2}). This came about accidentally. As Michael Berry might remember, I have been 
trying to work on various aspects of the theory of the fractional quantum Hall effect (QHE) for many years. 
At some point, I got interested in the seemingly purely academic question whether there are higher-dimensional 
cousins of the QHE; see \cite{ACF, FP}. Our answer to this question has turned out to be of interest 
also to people working on quantum optics and cold-atom physics, who then built on our work; 
see, e.g., \cite{Zilber}. Ruth Durrer kindly drew my attention to possible applications of our 
ideas to the problem of the origin of \textit{cosmic magnetic fields}, as envisioned in \cite{TW}. 
This has become quite a successful thread of our esearch; see \cite{ACF, FP, BFR1, BFR2}.
I also became interested in the \textit{mean-field limit} of quantum-mechanical many-body systems, originally 
studied by my PhD advisor Klaus Hepp (see \cite{Hepp, FKS}), and in solitary wave solutions of the limiting 
(mean-field) non-linear evolution equations, which have applications in studies of gravitational instabilities of boson- and 
neutron stars \cite{FL-1, FL-2} and in some models of axionic Dark Matter \cite{Duffy, Witten} (see also \cite{Fr1}).

More recently, I got interested in the problem of Dark Energy and of possible common origins of Dark Energy 
and Dark Matter and of the Matter-Antimatter asymmetry (MAA) in the \mbox{Universe --} of course, far too late to 
come up with really original proposals. Our efforts turned out to be related to ones earlier described 
in \cite{Wetter, LM, Wetterich, Peebles, Ratra}; see also \cite{JP, Copeland} for reviews. 
Our goal was to come up with a unified model accounting for DM, DE and MAA, as described in
\cite{BCFN, BFN, BF}. In the present paper, I present a brief survey of how far we got in reaching
that goal and constructing some models of the ``Dark Sector'' (leaving aside some of the work 
that is still going on and that I don't fully understand, yet). \\

\textit{Acknowledgements:} I thank my collaborators on topics related to ones featured in this paper for their 
valuable help; and, in particular, A.~Alekseev, H.~Bernardo, O.~Boyarsky, R.~Brandenberger, V.~Cheianov, 
E.~Lenzmann, B.~Pedrini and O.~Ruchayskiy. I also thank A.~H.~Chamseddine, R.~Durrer and N.~Straumann 
for instructive discussions on various matters not unrelated to ones discussed in this paper.

\subsection{A List of Puzzles in Cosmology}\label{puzzles}

To start with I present a list of some of the big puzzles encountered in the study of cosmology, 
accompanied by brief comments.
\begin{enumerate}
\item{Formation of ``classical'' structure from an initial quantum state of the Universe: What are \textit{``events''} in
a  quantum theory of the early Universe, and what role do they play in the emergence of structure? 
What is Dark Matter, and what is its role in the formation of ``classical'' structure in the Universe? --\\
Concerning an extension of quantum theory that can be expected to clarify what ``events'' are in Quantum 
Mechanics, why the occurrence of an event is always accompanied by ``information loss,'' and how events 
give rise to structure formation, I refer the reader to my work on the foundations of quantum mechanics; 
see \cite{Fr2, Fr3} and references given there.}
\item{Was there an era of \textit{Inflation}, and what would it explain? Is it natural? --\\
Standard wisdom has it that Inflation would explain the large-scale homogeneity, isotropy and spatial flatness of the observed
Universe. Observational indications of Inflation are found in the CMB: nearly scale-invariant fluctuations, acoustic peaks. 
See \cite{MC1, MC2}.}
\item{Why is the present expansion of the Universe \textit{accelerated}; what is Dark Energy? --\\
A speculative answer to this question forms the core of this review. I discuss a model that involves a scalar field
representing ``quintessence'' reminiscent of the one postulated in \cite{Wetterich}.}
\item{What is the origin of the observed Matter-Antimatter asymmetry in the Universe? --\\
The model introduced in the following features a pseudo-scalar axion field whose dynamics during 
a very early era in the evolution of the Universe could give rise to Matter-Antimatter asymmetry. }
\item{Why are there comparable amounts of Visible Matter, Dark Matter and Dark Energy in the Universe? 
Was this the case during earlier eras in the evolution of the Universe, and will it always be the case? -- \\
I have no idea of plausible, let alone correct answers.}
\item{What is the origin of the observed very weak and highly homogeneous cosmic magnetic fields that 
extend over intergalactic distances? --\\
Possible explanations are described in \cite{TW, FP, BFR1, BFR2}. I will not cover this topic in this review.}
\item{What are cosmological signs pointing to ``Physics beyond the Standard Model'' -- besides Dark Matter and Dark Energy? 
Are there clear indications of the existence of new degrees of freedom, such as \textit{WIMP's, axions, new 
scalar fields} and \textit{new gauge fields}, etc.~that can be extracted from observational data of cosmology? Are there any
signs in cosmology of the existence of extra dimensions? --\\
Well, these questions are related to those in items 2 through 6, and I won't add further comments.}
\end{enumerate}
 
The purpose of this paper is to review some features of a simple unified model of Dark Matter and Dark Energy \cite{BF}, thus
commenting on puzzles 1 and 3, and to suggest an application of the model towards resolving puzzle 4.

To prepare the ground for our discussion I proceed to recall some standard facts about the geometry 
of a homogeneous, isotropic Universe and the equations of state of matter, radiation and Dark Energy. 
For more in-depth information the reader will profit from consulting Michael Berry's book \cite{Berry}.

\section{Setting the Stage: The Geometry of a Homogeneous, Isotropic Universe}
From observational data on the CMB one infers that, up to an age of some $100'000$ years, before large-scale structures 
started to form, the Universe was remarkably \textit{homogenous} and \textit{isotropic}. This could be explained by
Inflation. Throughout the following I will treat the Universe as homogeneous and isotropic on very large 
distance scales ( $> 10^{7}$ pc \footnote{1 parsec (pc) $=  3.086 \times 10^{13}$ kilometres}, which are, 
however, much smaller than the optical radius, $\lessapprox10^{10} $ pc, of the Universe). Thus, 
the Universe may be thought to be foliated in space-like hypersurfaces, 
$\lbrace \Sigma_{t} \rbrace_{t \in \mathbb{R}}$, orthogonal to a time-like geodesic velocity field, 
$U= \frac{\partial}{\partial t}$, where $t$ is cosmological time. The induced metrics on the hypersurfaces 
$\Sigma_{t}, 0<t< \infty,$ are all proportional to one another. It follows that the Lorentzian
metric, $d\tau^{2}$, of the space-time of such an idealized Universe has the form
\begin{equation}\label{metric}
d\tau^{2}=dt^{2} - a^{2}(t) ds^{2},
\end{equation}
where $a(t)$ is a scale factor, and $ds^{2}$ is the metric of a 3D Riemannian manifold, 
$\Sigma$ (corresponding to $a=1$), of \textit{constant curvature}, $k$, with 
$$k=\frac{\varepsilon}{R^{2}}\,, \qquad \varepsilon = 0, \pm 1.$$
The parameter $R $ is the ``curvature radius'' of the 3D manifold $\Sigma$, and $\varepsilon$ 
has the following geometrical meaning:
\begin{itemize}
\item{$\varepsilon = -1$: the Universe is open and expanding for ever;}
\item{$\varepsilon = 0$: \,\,\, the Universe is spatially flat and expanding;}
\item{$\varepsilon = 1$: \,\,\, the Universe is spatially closed and will collapse.}
\end{itemize}
We plug the ansatz in \eqref{metric} into \textit{Einstein's field equations} of General Relativity, assuming that the 
energy-momentum tensor, $T=(T^{\mu}_{\,\nu}),$ is diagonal, 
$$T= \text{Diag}(\rho,-p,-p,-p),$$ 
and that appropriate equations of state hold that relate the \textit{energy density}, $\rho$, of the Universe 
to its \textit{pressure}, $p$.
Einstein's equations then reduce to the \textit{Friedmann equations} (see, e.g., \cite{Straumann})
\begin{equation} \label{Friedmann-1}
3H^{2} + 3 \frac{k}{a^{2}} = \kappa \rho + \Lambda_0,
\end{equation}
where the constant $\kappa$ is given by $\kappa= 8\pi G_{N}$, with $G_{N}$ Newton's gravitational constant, 
$H(t):=\frac{\dot{a}(t)}{a(t)}$ is the Hubble ``constant'', and $\Lambda_0$ is the cosmological constant 
(i.e., the coefficient of a term in the Einstein equations proportional to the metric tensor of space-time); 
and
\begin{equation} \label{Friedmann-2}
2\dot{H} - 2 \frac{k}{a^{2}} = -\kappa (\rho+p)
\end{equation}
Assuming that, initially, the Universe undergoes Inflation one may expect that $k=0$ (spatial flatness),
and we also set $\Lambda_0 =0$ (i.e., Dark Energy is assumed \textit{not} to be due to a cosmological constant)

By Eq. \eqref{Friedmann-1}, these assumptions are satisfied iff
$$\rho=\rho_{\text{crit.}}:=\frac{3}{\kappa} H^{2}\,.$$
One introduces a density parameter 
$$\Omega_{0}:= \frac{\rho}{\rho_{\text{crit.}}}.$$
Observational data suggest that 
$$\Omega_{0} \approx 1, $$ 
as would apparently be explained by Inflation! This implies that, besides Visible Matter (VM, $\approx 5\%$), Dark Matter (DM, 
$\approx 27\%$), there must also exist Dark Energy (DE, $\approx 68\%$), as confirmed by data from type IA 
supernovae (Perlmutter, Schmidt, Riess -- see, e.g., \cite{Perl}), from the CMB and from Baryon oscillations (BAO) 
in the power spectrum of matter.

Next, we recall the \textit{Equations of State} of Matter, Radiation and Dark Energy and recall the solutions 
of the Friedmann equations.

\begin{enumerate}
\item[(i)]{Visible and Dark Matter: \hspace{0.3cm}\, $p\approx 0$}
\item[(ii)]{Radiation: \hspace{2.9cm} $p=\frac{\rho}{3}$, \quad because\quad $T^{\mu}_{\,\mu}=0$ (conformal invariance)} 
\item[(iii)]{Dark Energy: \hspace{2.3cm}\, $p\approx -\rho$, \quad ($T^{\text{DE}}_{\mu \nu}\,\, \propto \,\,g_{\mu \nu}$\,!)}
\end{enumerate}
We solve the Friedmann equations with 
$$\rho + p = \delta \rho, \,\, 0<\delta <4/3,$$
 where $\delta=0$ corresponds 
to pure Dark Energy and $\delta=4/3$ to pure radiation. At present, Dark Energy contributes $\approx 68\%$ 
of the energy density of the Universe, and $\delta \approx 1/3$. The solution is given by
\begin{align}\label{Hubble}
a(t) = & \,\,a(t_0)\,\Big(\frac{t+\tau}{t_{0}+\tau}\Big)^{2/3\delta}, \nonumber \\
H(t) = & \,\,(2/3\delta) (t+\tau)^{-1}, \\
\rho(t) = & \,\,(4/3\kappa \delta) (t+\tau)^{-2} = const.\, a(t)^{-3\delta}\,, \nonumber
\end{align}
where $\tau$ is an arbitrary constant that we will set to 0.
\begin{itemize}
\item{For pure Radiation: \,$\delta=\frac{4}{3}$, hence  \,\,$\rho(t)\,\propto\, a(t)^{-4}\propto\, (t+\tau)^{-2}$ (radiation redshifts with increasing time).}
\item{For Visible Matter and/or Dark Matter only:\, $\delta=1, $ so that\, $\rho(t)\,\propto\, a(t)^{-3}\,\propto\, (t+\tau)^{-2}$.}
\item{For Dark Energy only:\, $\delta =0,$ hence, by Eq.~\eqref{Friedmann-2},\,  $H=const.,$\, \,and\, $\rho(t) = const.$}
\end{itemize}

If the Universe is in thermal equilibrium during the radiation-dominated phase (before recombination)
the \textit{Stefan-Boltzmann law} implies that 
\begin{equation} \label{temperature}
T(t)\,\,\propto\,\, \rho^{1/4}\,\, \propto\, \,\frac{1}{\sqrt{ t+\tau }}\,,
\end{equation}
and \,$T(t)=const.,$ for pure Dark Energy, i.e., $\delta=0.$

\section{A Simple Model of Dark Matter and Dark Energy}
In this section I introduce a model that might be expected to provide a unified mechanism explaining the 
presence of Dark Matter and of Dark Energy in the Universe and illuminating the origin of baryogenesis. 
At present, all such models are speculative, and the one considered in the following is no exception. It is
reasonable to require that the model involve as few degrees of freedom not already present in the Standard 
Model of particle physics as possible. These are the guiding principles for the choice of the model
sketched in the following (see \cite{BF}).

A conventional idea about Dark Matter is that it consists of particles called WIMPs ($=$weakly interacting massive particles, 
see, e.g., \cite{WIMP} for a review). Dark Energy is conventionally described by a small cosmological constant, $\Lambda_0$. 
Ordinary matter, WIMPS and a cosmological constant are the basic ingredients of $\Lambda$CDM models. 
However, the WIMP model of Dark Matter faces the problem that WIMP's have not been seen in any direct-detection experiments. Concerning Dark Energy, there is some evidence that a positive cosmological constant 
cannot be consistently introduced into current theories of quantum gravity (some work in this direction is quoted 
in \cite{BF}). It is therefore tempting to try to describe Dark Energy  by some new dynamical degrees of freedom; 
examples are those introduced in quintessence models; see \cite{Wetter} - \cite{Copeland}. 
Moreover, oscillating pseudo-scalar fields with a very small mass, typically axion fields, 
could serve as candidate degrees of freedom describing Dark Matter (see, e.g., \cite{axionDM} for a review). 

These considerations underly our proposal of a model of a complex scalar field, 
\begin{equation}\label{Z-field}
Z = e^{-(\sigma + i\theta)/f}\,,
\end{equation}
where the scalar field $\sigma$ is supposed to give rise to Dark Energy, and the pseudo-scalar (axion) field $\theta$ should 
account for Dark Matter. Furthermore, $f$ is a constant with the dimension of a mass (inverse length) rendering 
$(\sigma+i\theta)/f$ dimensionless. We will see that the model is only viable for values of $f$ large as 
compared to the Planck mass, $m_P$.

We set
$$\zeta_{\mu}:=Z^{-1}\partial_{\mu}Z= - f^{-1}(\partial_{\mu}\sigma + i\partial_{\mu} \theta), \quad \partial_{\mu}\equiv
\frac{\partial}{\partial x^{\mu}},\,\,\mu=0,...,3,\,\, x^{0}=t.$$
Let $U(r)$ be a non-negative polynomial in the variable $r$ with the property that $U(r=0)=0$. We will focus our attention on
the examples
\begin{equation}\label{self-int}
U(r)=U^{(2)}(r):= \Lambda r^{2}, \qquad \text{and} \qquad U(r)=U^{(4)}(r):= \frac{\Lambda}{r_{0}^{2}+\nu}\big[r^{2} (r- r_0)^{2}+ \nu\, r^{2}\big]\,,
\end{equation}
where $\Lambda$ is a constant with the dimension of a fourth power of mass, and $r_0 >0$ and $\nu\geq 0$ are 
dimensionless constants. We note that the two examples coincide near $r=0$.
Let $A$ be a possibly non-abelian gauge field (e.g., the weak $SU(2)$-gauge field), and let $\frak{j}$ be the 3-form dual to an anomalous current, $\mathfrak{J}^{\mu}$, (e.g., the baryon current) with the property that 
\begin{equation}\label{anomaly eq}
d\frak{j} = \frac{\alpha}{4\pi} \text{Tr}\big(F_{A}\wedge F_{A}\big) + \mathcal{O}(M)\,,
\end{equation}
where $F_{A}$ is the 2-form field strength of the gauge field $A$, $\alpha$ is a dimensionless coupling constant, 
and $M$ is a typical mass of matter fields appearing in the expression for $\mathfrak{J}^{\mu}$

Let $g_{\mu \nu}$ denote the metric tensor on space-time, and let $g$ denote its determinant.
We introduce an \textit{action functional}, $S$, for the field $Z$.
\begin{equation}\label{action-1}
S(\overline{Z}, Z):= \int \Big[\frac{f^{2}}{2}\, \overline{\zeta}_{\mu}\cdot \zeta^{\mu} - U(|Z|) - 
\lambda (\partial_{\mu} \text{Im} Z)\cdot \mathfrak{J}^{\mu} \Big] \sqrt{-g}\, d^{4}x\,,
\end{equation}
where $U$ is as in \eqref{self-int}, and $\lambda$ is a dimensionless coupling constant. When expressed in terms of the scalar field $\sigma$ and the axion field $\theta$ the action $S$ takes the form
\begin{equation}\label{action-2}
S(\sigma, \theta)= \int \Big[\frac{1}{2}(\partial_{\mu}\sigma\cdot \partial^{\mu}\sigma + \partial_{\mu}\theta\cdot 
\partial^{\mu} \theta) - U(e^{-\sigma/f}) + \lambda \partial_{\mu} \big(e^{-\sigma/f} \text{ sin}(\theta/f)\big)\cdot \mathfrak{J}^{\mu}
\Big] \sqrt{-g}\, d^{4}x\,, 
\end{equation}
with $U(e^{-\sigma/f}) \simeq \Lambda e^{-2 \sigma/f}$, as $\sigma \rightarrow + \infty$, for $U$ as in \eqref{self-int}.

After a phase transition (e.g., the electro-weak transition at $T_c \approx 160$ GeV), the gauge field $A$ is supposed
to acquire a mass. When intergrating out all massive degrees of freedom, with $g_{\mu \nu}$, $\sigma$ and $\theta$ 
treated as (classical) back ground fields, a low-energy theory results that has an effective
action of the form
\begin{equation}\label{eff action}
S_{eff}(\sigma, \theta)= \int \Big[\frac{1}{2}(\partial_{\mu}\sigma\cdot \partial^{\mu}\sigma + \partial_{\mu}\theta\cdot 
\partial^{\mu} \theta) - U(e^{-\sigma/f}) - V(\sigma, \theta)\Big] \sqrt{-g}\, d^{4}x\,,
\end{equation}
where $V(\sigma, \theta)\, (= \mathcal{O}(\theta^{2}), \text{ for } \theta \approx 0$) is a periodic function of the axion
field $\theta$ of the form
\begin{equation}\label{V-funct}
V(\sigma, \theta) \simeq \frac{1}{2}\mu^{4} e^{-2\sigma/f} \text{sin}^{2}(\theta/f)\,, \quad \text{ for small values of }\, 
e^{-\sigma/f}\,|\text{sin}(\theta/f)|,
\end{equation}
where $\mu$ is a constant with the dimension of mass.

\textit{Remark:} The action functionals in \eqref{action-1}, \eqref{action-2} and \eqref{eff action} do \textit{not} give 
rise to a renormalizable quantum field theory. One should therefore ask whether the effective field theory with action $S$ 
has an ``ultraviolet completion.'' It is tempting to think that superstring theory may yield a complete such theory.
This idea has been pursued in \cite{BBF}; but, at present, it is doubtful whether it is viable. 
In this paper, it won't be discussed any further. The functional $S$ will only be used as an action functional of a 
classical field theory describing low-energy (essentially classical) degrees of freedom governing the evolution 
of the space-time of the cosmos, with one-loop quantum corrections taken into account if necessary.

As usual, the field equations for the fields $\sigma$ and $\theta$ are derived by varying the action functional $S$ 
with respect to these fields. Since, in this note, we only explore the dynamics of an isotropic, homogeneous Universe,
we assume that $\sigma$ and $\theta$ only depend on cosmological time $t$. The equations of motion are then found 
to be
\begin{align}
\ddot{\sigma} + 3H\dot{\sigma} =& \,\Big[ \frac{2}{f} \Lambda+ \frac{\mu^{4}}{f}\text{sin}^{2}(\theta/f) \Big]\, e^{-2\sigma /f}\,, \label{sigma eq}\\
\ddot{\theta} + 3H\dot{\theta} =&-\frac{\mu^{4}}{f} \text{sin}(\theta /f) \text{cos}(\theta /f)\, e^{-2\sigma /f}\,, \label{theta eq}
\end{align}
where $H= \dot{a}/a$ is the Hubble constant; (possible further terms involving massive degrees of freedom are ignored).

We intend to explore consequences of the hypothesis that $\sigma$ plays the role of quintessence giving rise to  
Dark Energy, and oscillations of $\theta$ near a minimum of the interaction potential $V$ are a source of Dark Matter. 
Besides the fields $\sigma$ and $\theta$, radiation contributes to the pressure and the energy density of the early 
Universe. During some period after reheating, radiation is the dominant contribution to the energy density 
of the early Universe. As a consequence, the Hubble constant $H$ is positive, which, according to Eq.~\eqref{theta eq}, 
causes the oscillations of the axion field  $\theta$ to die out. When the amplitude of oscillations of $\theta$ becomes 
small, as is the case in the present era of evolution of the Universe, the first term on the right side of Eq.~\eqref{sigma eq} 
starts to dominate the evolution of the field $\sigma$.

If our model is to predict the observed energy densities of Dark Energy and Dark Matter in the Universe the amplitudes 
of the two terms on the right side of \eqref{sigma eq} must make comparable contributions at redshifts close to z = 2, 
just before Dark Energy starts to dominate. As in quintessence models, we assume that the initial value of $\sigma$ in the 
early Universe is small. The effective interaction potential $V$ for the axion field $\theta$ is assumed to
be generated at an early time corresponding to a temperature $T_c$ of a phase transition rendering the gauge field 
$A$ massive. Right after the time corresponding to $T_c$, the initial condition for $\theta$ is assumed to be
near a local maximum of $V$. We then consider three different periods in the evolution of the Universe: 
\begin{enumerate}
\item{The early era when $\theta$ is close to a local maximum of $V$, and radiation dominates;}
\item{an intermediate era when Dark Matter dominates over Dark Energy; and} 
\item{the late era when Dark Energy dominates.}
\end{enumerate}
\subsection{The Evolution of the Universe in the Intermediate and Late Era}
I proceed to discuss ideas about the evolution of the Universe during the intermediate 
and the late era, i.e., after the ``Dark Ages,'' (whence the title of this paper).
I begin by sketching what one might expect to happen during the intermediate era 2. 
Let us suppose that the self-interaction potential $U$ is given by $U(r)= U^{(4)}(r)$, 
with $\nu\approx 0$. Assuming that the evolution of the fields $\sigma$ and $\theta$ 
at the end of the early era 1 has started at suitable initial values, with $\sigma$ close to the local minimum, 
$\sigma = r_0$, of the potential $U$ and $\theta$ close to a local maximum of $V(r_0, \theta)$, then the field 
$\sigma$ is stuck in a \textit{metastable state} (of a possibly rather long life time), 
with $\sigma \approx r_0$, while the field $\theta$ slowly rolls 
down towards a local minimum of the effective potential $V(r_0, \theta)$, e.g., at $\theta =0$, and then starts 
to oscillate around that minimum with an initial amplitude of oscillation of $\mathcal{O}(f)$. 
The effective mass of the axion, i.e., of the field quanta of $\theta$, is given by 
$m_{axion}\,\approx \frac{\mu^{2}}{f} e^{-r_0/f}$. 
Since $U(\sigma =r_0) \approx 0$, the Dark Energy density very nearly vanishes during this 
intermediate period, while the oscillations of $\theta$ yield a large amount of massive Dark Matter with an 
equation of state $p\approx 0$. The energy density $\rho(t)$ of the Universe then decreases like $a(t)^{-3}$ 
(see Sect.~2). Since the Universe is expanding, with $H>0$, the oscillations of $\theta$ are 
damped (see Eq.~\eqref{theta eq}). Towards the end of era 2, a \textit{``cosmological 
wetting transition''}\footnote{For a discussion of the wetting transition see, e.g., \cite{wetting}. According to this
reference, the cosmological wetting transition may be expected to be continuous (of second order).}
sets in, as sketched in \cite{BFN}, and the field $\sigma$ tunnels out of the region near the 
local minimum of $U$ and evolves towards larger values. As a consequence, the value of 
$U$ becomes positive, meaning that a non-vanishing Dark-Energy density develops. The field $\sigma$ 
then starts to slowly roll down the exponentially decreasing slope of $U$ towards larger and larger values;
the second term on the right side of Eq.~\eqref{sigma eq} becomes negligibly small as compared to
the first term. The effective mass of the axion, 
$$m_{axion} \approx  \frac{\mu^{2}}{f} e^{-\sigma/f},$$
becomes \textit{time-dependent} and decreases towards smaller and smaller values. This implies that 
Dark-Matter lumps around galaxies tend to expand. After some time, the contribution of axions to the 
energy balance of the Universe becomes much smaller than the contribution of the degrees of freedom of 
$\sigma$; i.e., Dark Energy starts to dominate, and the late, Dark-Energy dominated period 3 in the evolution 
of the Universe sets in. (See \cite{BF} for a more quantitative discussion of era 2.)

The speed by which $\sigma$ rolls down the slope of $U$ towards larger and larger values is controlled 
by the parameter $f$. In order to get an equation of state compatible with observational data in the late period 3, 
$f$ must be proportional to the Planck mass, $m_{P}$, with a factor of proportionality that may be quite large,
as will be discussed next.
\subsection{Predictions Concerning the Late Era}
In the following, we focus our attention on the evolution of the Universe during the late era 3.
We make the ansatz that $\sigma$ only depends on cosmological time $t$ and that the metric of space-time satisfies the Friedmann equations; see \eqref{Friedmann-1} and \eqref{Friedmann-2}, with 
$$\rho+p = \delta \rho, \quad 0 < \delta < 4/3. $$
In the Dark-Energy dominated era 3, $\delta$ must be quite small, with $\delta \lessapprox \frac{1}{4}$. By Eq.~\eqref{Hubble}, we then have that $H(t) = (2/3 \delta) t^{-1}$ (w.l.o.g.~the constant $\tau$ in \eqref{Hubble} is set to $0$ in the following).
Neglecting the second term on the right side of \eqref{sigma eq} (or replacing $\text{sin}^{2}(\theta/f)$ on the right side of
\eqref{sigma eq} by its tiny mean value over a period of oscillation), the equation of motion for $\sigma$ is given by
\begin{equation}\label{sigma eq-2}
\ddot{\sigma}(t) + \frac{2}{\delta\, t} \dot{\sigma}(t) \approx \frac{2}{f} \Lambda_{eff}\, e^{-2\sigma/f}\,,
\end{equation}
where, in the following, the effective coupling constant $\Lambda_{eff}$ will be denoted again by $\Lambda$, and 
the dynamics of $\theta$ will now be neglected.\footnote{In estimating $\Lambda_{eff}$ one-loop quantum corrections of
the $\theta$-theory should be taken into account.}

The quantities $\rho$ and $p$ have to be calculated from the energy-momentum tensor, $T$, of the theory 
and must then be plugged into the Friedmann equations. The energy-momentum tensor in an isotropic, homogeneous
space-time has the form
$$T\equiv (T^{\mu}_{\,\,\nu})= \text{Diag}(\rho, -p, -p, -p).$$
The contribution of the field $\sigma$ to $T$ is calculated from the formula $T_{\mu\nu} = \frac{\delta S(\sigma, \theta=0)}{\delta g^{\mu\nu}}$\,, which implies
\begin{equation}\label{energy-momentum tensor}
T^{\mu}_{\,\,\nu}(x) = \frac{\partial \mathcal{L}(\sigma)}{\partial(\partial_{\mu}\sigma(x))}\cdot(\partial_{\nu}\sigma)(x) - \delta^{\mu}_{\,\,\nu} \mathcal{L}(\sigma)(x)\,,
\end{equation}
where $\mathcal{L}(\sigma) = \big(\frac{1}{2}\partial_{\mu}\sigma\cdot \partial^{\mu}\sigma - U(e^{-\sigma/f})\big)$. 
For simplicity, we may now choose $U$ to be given by $U=U^{(2)}$ (see \eqref{self-int}), 
which is justified for large values of $\sigma$. 
For our special ansatz, $\sigma=\sigma(t)$ (indep. of $\vec{x}$), this yields
\begin{equation}\label{energy-pressure}
\rho_{\sigma}= \frac{1}{2} \dot{\sigma}^{2} + \Lambda e^{-2\sigma/f}, \quad p_{\sigma}= \frac{1}{2}\dot{\sigma}^{2}-\Lambda 
e^{-2\sigma/f}.
\end{equation}
Setting $\rho= \rho_{\sigma} + \rho_{M},\, p= p_{\sigma} + p_{M}$,\, where $\rho_{M}$ is the energy density of matter and
$p_{M}\approx 0$ its pressure (the contribution of radiation can be neglected at late times), the Friedmann equations (with 
$k=0$ and a vanishing cosmological constant $\Lambda_0$) yield
\begin{equation}\label{Friedmann-3}
-\frac{2}{\kappa}\dot{H}=\rho+ p = \dot{\sigma}^{2}+ \rho_{M}= \delta \rho=\frac{3\delta}{\kappa} H^{2}.
\end{equation}

A special solution of Eq.~\eqref{sigma eq-2} is given by
\begin{equation}\label{solution}
\sigma(t) \equiv \sigma^{(0)}(t)=\sigma_{0}\,\ell n (\frac{t}{t_0}), \text{   with   }\,\sigma_{0}= f\,\, \text{and }\, t_{0}= 
\sqrt{\frac{2-\delta}{2\delta \, \Lambda}}\,f\,.
\end{equation}
For this solution, we have that
\begin{equation}\label{delta}
\rho_{\sigma}(t) + p_{\sigma}(t) = \delta \rho_{\sigma}(t), \quad \forall \delta\,,
\end{equation}
with
$$ \rho_{\sigma}(t)=\frac{f^{2}}{\delta}t^{-2}, \quad p_{\sigma}(t)=(1-\frac{1}{\delta})f^{2}t^{-2}.$$
Thus, the Friedmann equations are \textit{solved}, provided
\begin{equation}\label{solu}
\rho_{M}, p_{M} =0, \, \text{   and   }\,\,f^{2}= \frac{4}{3\delta \kappa}
\end{equation}
Tantalizingly, $f=4\kappa^{-1/2} = 4m_{P}$, for $\delta=\frac{1}{12}$\,.\\

\noindent
\textit{Remarks:}
\begin{itemize}
\item[I.]{Expression \eqref{solu} for $f^{2}$ suggests that the field $\sigma$ is a \textit{gravitational degree of freedom}. 
In work with \textit{Chamseddine} and \textit{Grandjean} \cite{CFG}, an exponential self-interaction potential 
for the field $\sigma$ has been obtained. Moreover, it has been suggested that $exp[-(2\sigma/f)]$ is related 
to the scale of an extra dimension (chosen to be discrete in \cite{CFG}), and that $f =\mathcal{O}(\kappa^{-1/2})$ is 
a consequence of deriving the action functional for $\sigma$ from a \textit{higher-dimensional Einstein-Hilbert action} 
by ``dimensional reduction''. Actually, a very similar idea had previously been proposed in \cite{Wetter}. 
Recently, it has been taken up again in \cite{BBF} in the context of superstring theory.}
\item[II.]{One may expect that, as time $t\rightarrow \infty$, when matter and radiation become negligible, the solution $\sigma^{(0)}$ is an \textit{``attractor''} in solution-space. This expectation is supported by the following result.}
\end{itemize}

\noindent
\textit{Theorem:} General solutions, $\sigma(t)$, of Eq.~\eqref{sigma eq-2} approach $\sigma^{(0)}(t)$, as $t \rightarrow \infty$.\\

\noindent
\textit{Linear Stability:}\\
Inserting the ansatz $\sigma(t):= \sigma^{(0)}(t) + \sigma^{(1)}(t),$ with $\sigma^{(1)}(t) \ll \sigma^{(0)}(t),$ for large $t$, into 
\eqref{sigma eq-2} and linearizing in $\sigma^{(1)}$, we find that
$$ \sigma^{(1)}(t)\,\, \propto\,\, t^{-\beta}, \quad \text{with }\,\,\beta = \gamma \pm \sqrt{\gamma^{2} - 8\gamma}, \quad 
\gamma:= \frac{2}{\delta} -1>\frac{1}{2}.$$
Note that $\text{Re}\, \beta \geq 0, \forall \delta \leq \frac{4}{3}$, with $\beta >0, \, \text{ if } \delta \leq \frac{2}{9}$;
hence $\sigma^{(1)}(t) \searrow 0$, as $t\rightarrow \infty$, for $\delta$ small enough. \\ 
If $\gamma < 8$, i.e.\, $\delta > \frac{2}{9},$ then $\text{Im}\,\beta \not= 0$, hence $ \sigma^{(1)}$ describes 
\textit{ oscillations} with a tiny time-dependent mass $\propto f (t_{0}/t)^{2}$) that die out like 
$t^{\frac{1}{2}-\delta^{-1}}$. \\

\noindent
\textit{Non-Linear Stability:}\\
$$\rho_{\sigma}= \frac{1}{2}\dot{\sigma}^{2} + \Lambda e^{-2\sigma/f}$$
is a \textit{Lyapunov functional} that decreases in time on solutions of \eqref{sigma eq-2}. \\ 
\textit{All} solutions of \eqref{sigma eq-2} are bounded above by $\ell n(\frac{t}{t_{*}})$, for some $t_{*}$.\\
\subsection{Observational Constraints on Model Parameters}
In order for the model discussed in this paper to be compatible with observational data some constraints
need to be imposed on the values of the parameters $f, \Lambda$, and $\mu$ in the action functional 
introduced in \eqref{eff action} and \eqref{V-funct} and on the value of the transition temperature $T_c$ 
below which the gauge field $A$ becomes massive and the effective potential $V$ for the axion field 
$\theta$ is generated.
Let $t_0$ denote the present time. As argued above, for the field $\sigma$ to qualify as quintessence, 
predicting an acceptable equation of state typical of the Dark-Energy era, we must require that 
$f = \mathcal{O}(m_P)$.

In the Dark-Energy era, the first term on the right side of Eq.~\eqref{sigma eq} must dominate over the second term.
For this to happen one must demand that
$$\Lambda e^{-2\sigma(t_0)/f} \approx T_0^{4} z_{eq}\,,$$
where $T_0$ is the present  temperature of the CMB, and $z_{eq}$ is the redshift at the time of equal matter 
and radiation (besides imposing a bound on the parameter $\mu$); see \cite{BF}. The Stefan-Boltzmann law 
is used in this relation.

The parameter $\mu$ is constrained by the requirement that the present amplitude, $\mathcal{A}(t_0)$,
of oscillations of the axion field $\theta$ be compatible with the presently observed contribution of Dark 
Matter to the energy density of the Universe. This amounts to imposing the condition that
$$\mu^{4}\,\big(\mathcal{A}(t_0)/f\big)^{2} e^{-2 \sigma(t_0)/f} = \mathcal{O}(T_{0}^{4} z_{eq})\,.$$
Assuming that $\sigma(t_0) = \mathcal{O}(f)$, we may replace $e^{-2 \sigma(t_0)/f}$ by a constant,
$\mathcal{O}(1)$, smaller than 1 in the two conditions just stated. Using a straightforward estimate on 
$\mathcal{A}(t_0)$, one then finds that
\begin{equation}\label{param values}
\Lambda = \mathcal{O}(T_{0}^{4} z_{eq})\,, \qquad \mu^{4} \Big(\frac{T_0}{T_c} \Big)^{3} = \mathcal{O}(T_{0}^{4} z_{eq})\,.
\end{equation}
As Eq.~\eqref{theta eq} shows, the value of the parameter $\mu$, along with the value of $\sigma(t_0)/f$, determines
the mass, $m_{axion}$, of the axion, which, in our model, is the mass of the Dark-Matter particle. Setting $e^{-2\sigma(t_0)/f}$
to a constant of $\mathcal{O}(1)$, we find that 
$$m_{axion} = \mathcal{O}(\mu^{2}/f)\,.$$
These considerations yield
\begin{equation}\label{axion mass}
\mu^{2} =\mathcal{O}(m_{axion} \cdot m_{P}), \qquad T_c = \mathcal{O}(m_{axion})\times 10^{20}\,.
\end{equation}
To end up with a realistic value for $T_c$ one must assume that the present value of the axion
mass is $m_{axion} = \mathcal{O}(10^{-20} - 10^{-12})$ eV, i.e., the axion of the Dark-Energy era must be very light.
(Clearly, the order of magnitude of $m_{axion}$ depends on the nature of the phase transition happening at the temperature $T_c$, which we presently leave open.)

The idea that a very light axion, as considered above, may be a candidate particle for so-called 
\textit{``fuzzy Dark Matter''} has been rather widely discussed in the literature. For a recent 
analysis of this idea see \cite{Witten} and references given there.\footnote{Unfortunately, the analysis in \cite{Witten}
involves approximations that appear to be illegitimate, given results in \cite{FKS, FL-1, FL-2}.}  
So far, no direct detection of Dark Matter particles has been reported.
This fact may favour models involving very light Dark-Matter particles such as the one considered in this paper. 
Furthermore, for axion masses like those in \eqref{axion mass}, the axion as a source of Dark Matter is unlikely 
to form ``small-scale'' structure in the Universe, another welcome feature of our model.\\

Returning to Eqs.~\eqref{solution} and \eqref{solu} and recalling the Theorem stated after Remark II, one
concludes that the following scenario for the late-time evolution of the Universe is plausible:

Constraining the model-parameter values such that an acceptable equation of state holds in the late-time 
(Dark-Energy) era one is led to expect that, at very late times in the evolution of the Universe, the degrees 
of freedom of radiation, Dark Matter and visible matter can be neglected and the behavior of ``quintessence'' 
is described (at least qualitatively) by the solution $\sigma(t) \equiv \sigma^{(0)}(t)=\\ =\sigma_{0}\,\ell n (\frac{t}{t_0})$ 
displayed in Eq.\eqref{solution} of the equation of motion \eqref{sigma eq-2} (with $\rho + p \ll \rho$).

On the basis of our model it is quite safe to predict that the large-time fate of the Universe, dominated
by Dark Energy, will be very boring; i.e., \textit{the Universe must be expected to end in a Dark Age!}\\

\subsection{Some Comments on Baryogenesis}
We conclude this section with a few sketchy comments on physics in the very early era (era 1), in particular on 
\textit{baryogenesis} and Matter-Antimatter Asymmetry (MAA).
We return to Eqs.~\eqref{anomaly eq} and \eqref{action-1}, with $\mathfrak{J}^{\mu}$ the baryon current of the 
standard model. As argued in \cite{Shaposhnikov, Riotto} and in \cite{Anber} (see also \cite{FP, BFR1} for more 
recent accounts of related ideas), the coupling of the anomalous baryon current $\mathfrak{J}^{\mu}$ to the 
gradient of a pseudo-scalar field, as in \eqref{action-1} and \eqref{action-2}, can trigger the growth of baryon 
number and MAA in the early Universe.\footnote{For this to happen the phase transition at the temperature $T_c$ 
may have to be discontinuous, i.e., of first-order.} 
To see how this may work we first notice that the current
\begin{align}\label{cons current}
\widetilde{\mathfrak{J}}^{\mu} := \,&\mathfrak{J}^{\mu} - \frac{\alpha}{2\pi}CS^{\mu}(A), \quad \text{ where }\nonumber\\
CS^{\mu}(A) := \,& \varepsilon^{\mu \nu \kappa \rho}\,\text{Tr}(A_{\nu} F_{\kappa \rho} + \frac{2}{3} A_{\nu}
A_{\kappa} A_{\rho})
\end{align}
is the dual of the Chern-Simons 3-form, is \textit{conserved}, but \textit{not} gauge-invariant. However, the charge
$$\mathfrak{Q}:= \int_{\sigma_t} \widetilde{\mathfrak{j}}\,,$$
where $\widetilde{\mathfrak{j}}$ is the 3-form dual to the current $\widetilde{\mathfrak{J}}^{\mu}$, is 
a \textit{gauge-invariant, conserved} charge. (To simplify the following discussion we treat 
the gauge field $A$ as a classical external field. However, 
one can treat this field as quantized and introduce appropriate expectation values 
at the right places; see, e.g., \cite{Fr1}.) Let us imagine that, at temperatures larger than $T_c$ 
(before the gauge field $A$ becomes massive), the Chern-Simons 3-form of the gauge field $A$ 
is non-zero, with 
\begin{equation}\label{helicity}
CS^{0}(A)\equiv \mathfrak{h} \not= 0
\end{equation}
a non-zero helicity density that depends on time but is approximately constant in space. After the phase transition 
at the time $t_c$ corresponding to the temperature $T_c$, the gauge field $A$ becomes massive, and $CS^{0}(A)$ 
vanishes thereafter. Conservation of $\mathfrak{Q}$ then implies that the density $\mathfrak{j}^{0}$ changes by 
an average amount given by $\mathfrak{h}$; i.e., a non-zero baryon density is generated during 
this phase transition, assuming that $\mathfrak{h}$ was non-zero in the high-temperature phase.

The growth of a non-vanishing helicity density $\mathfrak{h}$ at times earlier than $t_c$ can be the result of
processes triggered by the last term, $-\lambda \partial_{\mu} \big(e^{-\sigma/f} \text{ sin}(\theta/f)\big)\cdot \mathfrak{J}^{\mu}$,
within the bracket under the integral of the action functional in Eq.~\eqref{action-2}, which, after an integration by parts, corresponds to
$$\frac{\lambda \alpha}{4\pi} e^{-\sigma/f}\text{sin}(\theta/f)\, \text{Tr}(F_A\wedge F_A) + \mathcal{O}(M)\,$$
see Eq.~\eqref{anomaly eq}.
For such processes to be effective, one assumes that there exists an era during which the field 
$ e^{-\sigma/f} \text{sin}(\theta/f)$ slowly rolls from an initial value $\approx \lambda e^{-r_0/f}$ 
towards values close to 0. 
That slow roll of a pseudo-scalar axion field leads to a (spatially rather uniform) non-zero helicity density $\mathfrak{h}$
has been shown in \cite{Shaposhnikov, Riotto}. The main arguments underlying this claim have been 
recalled in \cite{BF}. Related arguments have been used to propose a possible mechanism that may 
explain the presence of tiny, highly uniform primordial magnetic fields in the Universe extending over
intergalactic disctances; (see \cite{TW, Fr1,FP, BFR1} and references given there).\footnote{It should be mentioned
that the axion field involved in the discussion just presented may differ from the field $\theta$ introduced earlier.} 
For more details the reader is referred to the literature quoted above.

\section{Concluding Remarks}
To an outsider like myself, theoretical cosmology -- in contrast to observational, phenomenological 
and computational cosmology -- does not look like a firmly established science, yet. The following 
features of theoretical cosmology appear to point to serious difficulties in our understanding of key problems 
that will have to be overcome in the future.
\begin{enumerate}
\item{The coupled partial differential equations describing the evolution of radiation, visible matter, Dark Matter, 
Dark Energy and the geometry of space-time are highly non-linear. They may be expected to exhibit instabilities, 
in particular gravitational instabilities, that we do not know how to treat properly, yet.\footnote{Incidentally, 
this is a proviso against the treatment of fuzzy dark matter presented in \cite{Witten}.}}
\item{In every full-fledged analysis of the evolution of the Universe, sooner or later, one faces the necessity to treat all degrees 
of freedom of radiation and matter quantum-mechanicallly; but, \textit{faute de mieux}, all gravitational degrees of freedom
(the metric $g$, the Dark-Energy field $\sigma, \dots$) are treated classically. This leads to logical inconsistencies. 
While there may be various self-consistent ways (semi-classical approximations) to defuse this fundamental problem, 
it is deeply disturbing that we still do not know how to combine Quantum Theory with a Relativistic Theory of Gravitation}
in a mathematically consistent-looking theory.
\item{On the positive side, a case can be made for the existence of additional gravitational degrees of freedom (in
this paper in the form of the field $Z=e^{-(\sigma+i\theta)/f}$) accounting for Dark Matter and Dark Energy. This is 
bound to inspire thoughts on ``physics beyond the Standard Model,'' which will hopefully bear fruit in the future.}
\item{The form of the effective potential, $\propto\, e^{-2\sigma/f}$, of the field $\sigma$, which gives rise to 
Dark Energy, appears to emerge from different scenarios involving \textit{extra dimensions}, in particular from 
superstring theory (see \cite{BBF}). Although we do not know any quantitatively satisfactory derivation of this
potential, yet, we may feel encouraged to take theories with extra dimensions seriously.}
\end{enumerate}
I should stress that there are plenty of competing recent ideas about the nature of Dark Matter and 
Dark Energy. As one example I mention an intriguing proposal made in \cite{CM-1, CM-2}.
\begin{center}
---
\end{center}

To conclude, I return to the theme alluded to at the beginning of this paper with a comment concerning the 
``Dark Age'' that may loom over humanity. I quote the eminent mathematician Alexander Grothendieck, 
who, more than fifty years ago, said (see \cite{Grothendieck}):\footnote{I trust that Grothendieck's text can be understood
by people who are not fluent in French.}

\textit{... depuis fin juillet 1970 je consacre la plus grande partie de mon temps en militant pour le mouvement 
\textit{``Survivre''}, fond\'{e} en juillet  \`{a} Montr\'{e}al. Son but est la lutte pour la survie de l'esp\`{e}ce humaine, 
et m\^{e}me de la vie tout court, menac\'{e}e par le d\'{e}s\'equilibre \'ecologique croissant caus\'e par une utilisation 
indiscrimin\'ee de la science et de la technologie et par des m\'ecanismes sociaux suicidaires, et menac\'ee 
\'egalement par des conflits militaires li\'es} [\`a la politique d'h\'eg\'emonie des grandes puissances et] \textit{\`a la prolif\'eration des appareils militaires et des industries 
d'armements. ...}\\

Sadly, we have apparently not learned much, if anything, during the past fifty years!

\begin{center}
-----
\end{center}

\bigskip

\noindent
J\"urg Fr\"ohlich, ETH Zurich, Department of Physics, HIT K42.3, CH-8093 Zurich, Switzerland;\\\href{mailto:juerg@phys.ethz.ch}{juerg@phys.ethz.ch}

\end{document}